# An integral gated mode single photon detector at telecom wavelengths

Zhengjun Wei, Peng Zhou, Xiaobao Liu, Jindong Wang, Changjun Liao, Jianping Guo, Rihao Li, and Songhao Liu

School for Information and Optoelectronic Science and Engineering,
South China Normal University, Guangzhou 510631, China

*Abstract*—**We demonstrate an integral gated mode single photon detector at telecom wavelengths. The charge number of an avalanche pulse rather than the peak current is monitored for single-photon detection. The transient spikes in conventional gated mode operation are canceled completely by integrating, which enables one to improve the performance of single photon detector greatly with the same avalanche photodiode. This method has achieved a detection efficiency (*DE*) of 29.9% at the dark count probability per gate equal to $5.57 \times 10^{-6}$ /gate ($1.11 \times 10^{-6}$ /ns) at 1550nm.**

The single photon detectors (SPD) used at telecom wavelengths have become a focus with the boom of quantum cryptography technique [1]. At present, the best choice is the InGaAs/InP avalanche photodiode (APD) with separate absorption multiplication (SAM) at telecom wavelengths. The APD is operated above its breakdown voltage, which is called Geiger mode operation [2], so that a photon induced avalanche can grow into a macroscopic pulse. If the amplitude of a avalanche pulse exceeds the threshold of the pulse height discriminator, it will trigger a count and the smaller pulses will be rejected with the system noise by the discriminator.

To improve the performance of the SPDs, one way is developing the design and manufacturing process technology of the sensors, the other is developing the operation mode and the bias circuit. The gated mode single photon detector (GSPD) is the most effective method to decrease the dark count probability and afterpulse probability by using the gate pulses synchronized with the arrival of photons [3]. Although this method works well, the use of short gate pulses leads following drawback. The gate pulses coupled to the load resistor through the APD's junction capacitor, as shown in Fig.2.curve (2), produce two transient spikes that obscure the photon-induced avalanche pulses. High threshold voltage of the pulse height discriminator is necessary to avoid the transient spikes triggering a false count. Generally, in a pulse height distribution of the APD output signals for single photons, the probability that lower-height pulses are generated is greater than that for higher-height pulses [4]. Therefore, High threshold voltage decreases the detection efficiency of the SPD.

The transient spikes are the main problem that degrades the performance of a GSPD when the APD is certain. Much of the recent works on SPDs at telecom wavelengths have focused on solving this problem. These works include the coaxial cable reflection lines scheme [5,6], the two similar APDs balanced scheme [7], the capacitor balanced scheme [8], and the discharge pulse counting scheme [9]. All these techniques are complex and need to be elaborately regulated. Even so, they can not cancel the spikes completely or have drawbacks. Provided that the delay lines are matched and respond linearly, the proposal of using coaxial cable reflection lines is the only method that can cancel the spikes independent of pulse height, duration, or shape. Although this scheme employs two gate pulses during one detection cycle which leads to doubled dark counting probability and halved detection frequency, benefited from transient spikes cancellation, they still achieved the best performance to date. New proposal is needed to solve the transient spikes problem and avoid introducing new drawback.

In this paper, an integral gated mode single photon detector (IGSPD) is introduced. In this operation mode, the charge number of the avalanche pulse rather than the peak current of the APD is monitored for single photon detection. It was demonstrated at telecom wavelengths and the experimental results are compared with the coaxial cable reflection line scheme using the same type of APD.

The schematic of the integral gated single photon detector is shown in Fig.1. The InGaAs/InP APD is biased by an AC coupled bias tee and an integral capacitor is employed in the single photon detector instead of the 50Ω load resistor. The integral capacitor and the charge amplifier compose an integrator. The expression of the voltage across the integral capacitor can be derived as

$$v_C(t) = v_C(t_0) e^{\frac{-(t-t_0)(C_L+C_{APD})}{R_i C_L C_{APD}}}$$
$$+ \frac{(C_L + C_{APD})}{R_i C_L C_{APD}} e^{\frac{-t(C_L+C_{APD})}{R_i C_L C_{APD}}} \int_{t_0}^{t} v_p(\xi) e^{\frac{\xi(C_L+C_{APD})}{R_i C_L C_{APD}}} d\xi$$
$$+ \frac{(C_L + C_{APD})}{R_i C_L C_{APD}} e^{\frac{-t(C_L+C_{APD})}{R_i C_L C_{APD}}} \int_{t_0}^{t} [iR_i + \int_{t_0}^{t} i(\tau)d\tau] e^{\frac{\xi(C_L+C_{APD})}{R_i C_L C_{APD}}} d\xi$$

(1)

where $v_p(t)$ is the voltage of the gate pulse, $C_L$ is the capacitance of the load capacitor, $C_{APD}$ is the APD's junction capacitance, and $i(\tau)$ is the body current of the APD including dark current and photon-induced current, $R_i$ is the internal resistance of the gate pulse generator. The first term is the zero input response, the second is the response to the gate pulse, and the third is contributed by the body current of the APD. This voltage signal is amplified for discriminating by a charge amplifier. The integral capacitor is discharged by an electronic switch in order to reset the circuit for the next detection. A low pass filter and a digital averager are used to compress the system noise further.

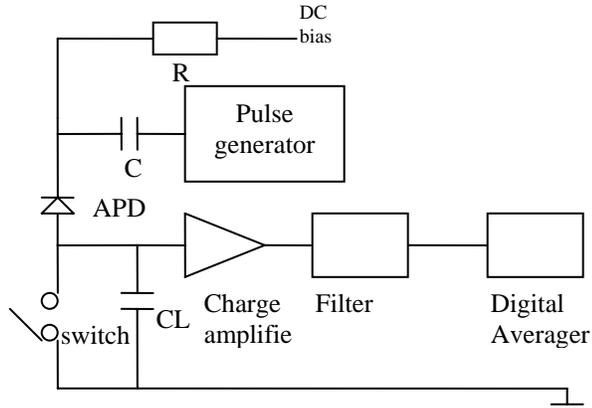

Fig.1. Schematic of the integral gated mode single photon detector

The range of integration can be set by controlling the open and close time of the switch. If the range of integration is set from A to B, including the gate pulse as shown in Fig.2.curve (3), the first and second term of equation (1) become zero. It is because that the APD's junction capacitor is in series with the integral capacitor and the two capacitors will be charged by the rising edge of the gate pulse and discharged by the falling edge simultaneously. Because the bias voltage is constant at the DC voltage before and after the gate pulse, the total charges which the gate pulse injects into the integral capacitor are zero. Therefore the output voltage of the charge amplifier only corresponds to the number of avalanche charges as shown in Fig.2.curve (3) after the integration. The transient spikes can be eliminates completely independent of pulse height, duration, shape, or type of APD even if without any regulation.

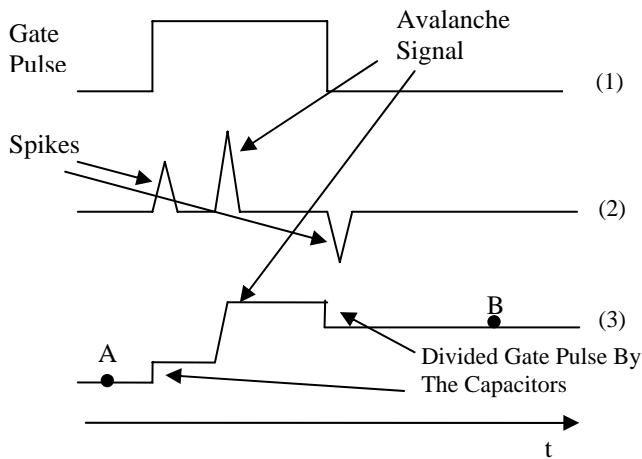

Fig.2. Comparison between the output of the conventional gated mode single photon detector and the integral gated mode single photon detector. The curve (1) is the steep gate pulse, the curve (2) is the output of the normal GSPD, and the curve (3) is the output of the IGSPD.

Once transient spikes have been canceled completely, the system noise of the GSPD can not be ignored any more. The full width at half maximum (FWHM) of the avalanche pulse is typically 500ps. This requires the amplifier whose bandwidth exceed 1GHz. The typically root-mean-square value of equivalent input noise is 126μV for a 1GHz, high performance amplifier. Furthermore, considering the thermal noise of the 50 Ω load resistor, the threshold of the discriminator must exceed 1664μV to avoid false count caused by the noise at 298K. Consider the gain distribution of APD, such high threshold voltage degrades the performance of the SPD system, while the avalanche pulse has peak amplitude of –5 mV [3, 5].

Compressing the bandwidth is a common technique to increase the sinal-to-noise ratio(SNR). By transforming the ultra fast avalanche current into static charges holding in the integral capacitor, the integral operation mode makes it possible for us to use low pass filter and digital averager. The system noise can be one order of magnitude lower than the conventional GSPDs by employing a 10MHz bandwidth lowpass filter. The threshold of the discriminator can be reduced to 83.2uV to allow more avalanche trigger a count.

The gate frequency of IGSPD is determined by the response of the charge amplifier (AD8067) and the switch (ADG751). The –3 dB bandwidth of AD8067 is 54 MHz at gain = +10. The turn on time ($t_{ON}$) and turn off time($t_{OFF}$) of ADG751are typically 9 ns and 3 ns, so the switching rate is 83MHz. There are many devices that faster than the chosen and the reported fastest SPD was operated at 14MHz, therefore the operation frequency of SPDs will not be limited by our scheme.

A picosecond diode laser (Sepia PDL808, Picoquant) emits a sequence of light pulses, each having a width of 50 ps at 1545 nm. Single photons are obtained by attenuating laser pulses with a variable attenuator. The APD used in the experiment was selected is ETX40 from JDS Uniphase. The APD is temperature stabilized at 224 K with a residual fluctuation of 0.1 K. The gate pulses of 5.13ns FWHM and 4.4Vpeak to peak value are applied to the APD after they combined with DC bias of 43.1V. The breakdown voltage of this APD is 46.6V at 224K. So the excess reverse voltage ($v_E$) above the breakdown voltage is 0.9V. In order to study the SNR of the IGSPD more accurately, the master clock of the IGSPD is kept at 100kHz to eliminate the afterpulse effect.

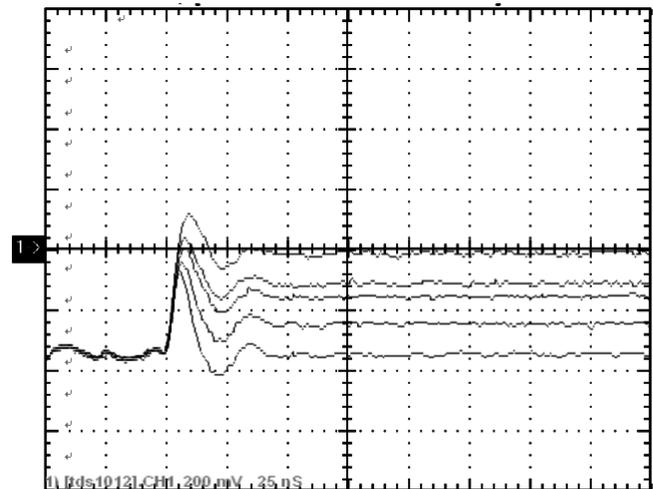

Fig.3. The output waveforms of the charge amplifier responding with single photon incidence monitored by an oscilloscope TDS1012.

The transient spikes cancellations we achieve are shown in

Fig.3. The bottom trace corresponds to the output waveform of the charge amplifier without avalanche and the upper traces correspond to the avalanche pulses superimposing upon the gate pulses divided by the capacitors. The traces become flat with different amplitude after the gate pulses become zero. Therefore the sampling points can be set at the peak point or after the gate pulse. The different levels correspond to the different random avalanche gains of the APD[4].

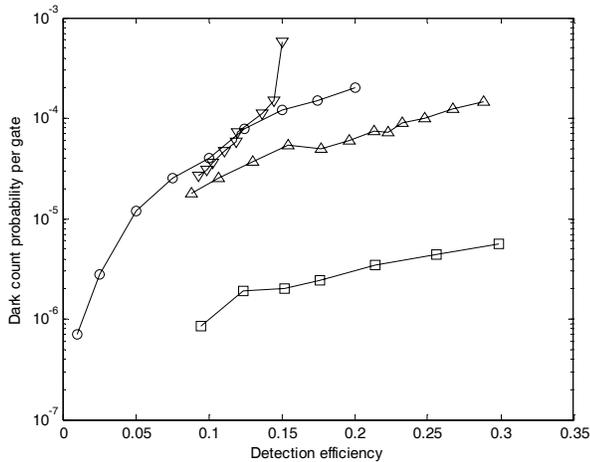

Fig.4. Dark count probability per gate versus $DE$ for: ($\nabla$) a resistance load gate mode single photon detector without spikes canceling; (O) a resistance load gate mode single photon detector with spikes canceling; ($\triangle$) and ($\square$) the integral gate mode single photon detector at $V_E = 1.3V$ and 0.9 ,respectively.

The single photon detecting performance of the IGSPD is shown in Fig.4 ($\triangle$) and ($\square$) at $V_E = 1.3V$ and $0.9V$ ,respectively. For comparing with the IGSPD, an experiment of a conventional resistance loading GSPD without transient spikes canceling circuit has been done with the same APD. The operation condition is same as the IGSPD, except the $V_E$ has been increased to 1.3V to obtain the best performance. The experimental results are shown in Fig.4 ($\nabla$). Another experiment data using the same type of APD with transient spikes canceling circuit at 100KHz, 223K and 1550nm wavelengths is also cited in Fig.4 (O) for reference [10].

As shown in Fig.4, the performance of the resistance load GSPDs with and without spikes canceling circuit is similar when the detection efficiency is under 14%. However, the dark count probability of the SPD without transient spikes canceling circuit increases rapidly after the *DE* exceeding 14%, because the discriminating threshold voltage is near the voltage of the positive transient spike. The GSPD with spikes canceling circuit can get rid of this problem, but the whole dark counting probability is bigger than the IGSPD's for high system noise and the doubled dark counting probability.

*DE* is the overall probability of registering a count if a photon arrives at the detector, and includes fiber coupling loss, APD optical coupling efficiency and intrinsic quantum efficiency, and the efficiency with which the signal processing electronics respond to photon signals from the APD.

The dark count can be randomly triggered by carriers generated in thermal, tunneling or trapping processes taking place in the junction. The occurrence of thermally generated carriers can be reduced by cooling the APD. At low temperature, dark counts are thus dominated by carriers generated by band to band tunneling and trapped charges. Decreasing the excess bias voltage can reduce the occurrence of dark counts.

Depending on the powerful charges gathering capability of the integral capacitor and the low noise system of the IGSPD, the threshold voltage of the discriminator can be set at 20mV. Considering the amplifier gain of +10 and a 33pF integral capacitance, the avalanche pulses whose charge number exceeds 0.066pC can trigger a count. Therefore the excess bias voltage of the IGSPD is 0.4V lower than the conventional resistance loading GSPD and a single photon detection efficiency of 29.9% at dark count probability per gate $P_D = 5.57 \times 10^{-6} / gate$ or $1.11 \times 10^{-7} / ns$ has been achieved. The improved results of the IGSPD indicates that because the relatively smaller avalanche pulses are rejected by the high threshold of the discriminator, the measured detection efficiency of the conventional GSPDs is much lower than the intrinsic quantum efficiency of the APD.

In conclusion, we have developed an integral gated mode single photon detector and demonstrated that it can efficiently improve the performance of the SPD with the same APD. Canceling the spikes and compressing bandwidth enabled us to reduce the threshold at the discriminators and thus decrease bias voltage, which reduces the dark count probability without sacrificing detection efficiency. The paper would help on design single photon detectors.